\documentclass[cameraready]{Interspeech}
\title{Generating Synthetic Doctor-Patient Conversations for Long-form Audio Summarization}

\author[affiliation={1}]{Yanis}{Labrak}
\author[affiliation={2}]{David}{Grünert}
\author[affiliation={1,4}]{Severin}{Baroudi}
\author[affiliation={3}]{Jiyun}{Chun}
\author[affiliation={5}]{Pawel}{Cyrta}
\author[affiliation={1}]{Sergio}{Burdisso}
\author[affiliation={6}]{Ahmed}{Hassoon}
\author[affiliation={7}]{David}{Liu}
\author[affiliation={8}]{Adam}{Rothschild}
\author[affiliation={9}]{Reed}{Van Deusen}
\author[affiliation={1}]{Petr}{Motlicek}
\author[affiliation={3}]{\\Andrew}{Perrault}
\author[affiliation={4,10}]{Ricard}{Marxer}
\author[affiliation={11,12}]{Thomas}{Schaaf}

\address{
    $^1$ Idiap Research Institute, Switzerland \hspace{3mm}
    $^2$ University of Zurich, Switzerland \\ %\hspace{3mm}
    $^3$ The Ohio State University, USA \hspace{3mm}
    $^4$ Université de Toulon, Aix Marseille Univ, LIS, CNRS, France \\
    $^5$ Stenograf, Poland \hspace{3mm}
    $^6$ Johns Hopkins University Bloomberg School of Public Health, USA \\ %\hspace{3mm}
    $^{7}$ Colorado School of Mines, USA \hspace{3mm}
    $^{8}$ Allegheny Health Network, USA \\%\hspace{3mm}
    $^{9}$ University of Pittsburgh Medical Center, USA \hspace{3mm}
    $^{10}$ ILLS, CNRS, France \\ %\hspace{3mm}
    $^{11}$ Solventum, USA \hspace{3mm}
    $^{12}$ Carnegie Mellon University, USA
}

\email{tschaaf@solventum.com}

\newcommand{\ci}[2]{\makecell{$#1$\\[1pt]{\footnotesize$\pm#2$}}}

\usepackage{comment}
\usepackage{multirow}
\usepackage{makecell}
\usepackage{caption}
\usepackage{subcaption}
\usepackage{siunitx}

\begin{document}

\maketitle

\begin{abstract}
Long-context audio reasoning is underserved in both training data and evaluation. Existing benchmarks target short-context tasks, and the open-ended generation tasks most relevant to long-context reasoning pose well-known challenges for automatic evaluation. We propose a synthetic data generation pipeline designed to serve both as a training resource and as a controlled evaluation environment, and instantiate it for first-visit doctor-patient conversations with SOAP note generation as the task. The pipeline has three stages---persona-driven dialogue generation, multi-speaker audio synthesis with overlap/pause modeling, room acoustics, and sound events, and LLM-based reference SOAP note production---built entirely on open-weight models. We release 8,800 synthetic conversations with 1.3k hours of corresponding audio and reference notes. Evaluating current open-weight systems, we find that cascaded approaches still substantially outperform end-to-end models.
\end{abstract}

\section{Introduction}
Toward the broader goal of human-level audio understanding, recent large audio language models (LALMs)~\cite{xu2025qwen3omnitechnicalreport,goel2025audioflamingo3advancing,Intelligence2025,li2025baichuanomni15technicalreport} have demonstrated impressive progress on benchmarks for audio processing and comprehension~\cite{ghosh2025audio,ahia2025blabbrutallylongaudio,chen2025audiollmsreallylisten,kumar2025mmauprochallengingcomprehensivebenchmark}. However, these benchmarks focus predominantly on short-context tasks. Our understanding of LALM performance on long-context audio reasoning ($>$ 5 minutes) remains limited, in part because LALMs capable of accepting long-context input only emerged in mid-2025, and in part because constructing meaningful benchmarks for such tasks is difficult.

Beyond data scarcity, evaluation itself is a bottleneck. Automatically constructed benchmarks risk covering only a narrow slice of audio understanding, and the tasks most relevant to long-context reasoning — such as summarization and note-taking — require open-ended generation, where multiple valid outputs exist for any given source, surface overlap metrics correlate weakly with human quality judgments~\cite{liu-etal-2016-evaluate}, and fluent outputs can hallucinate content not present in the source~\cite{maynez-etal-2020-faithfulness}. Rather than contributing another benchmark, we propose a synthetic data generation pipeline that can serve both as a source of training signal (via supervised fine-tuning or reinforcement learning with verifiable rewards) and as a controlled evaluation environment.

% \noindent We make the following contributions:
% \begin{itemize}
% \item A full synthetic pipeline---persona-based dialogue generation, audio synthesis featuring audio synthesis featuring overlap/pause modeling, sound event insertion, and room acoustic simulation and reference SOAP note production---built entirely with open-weight, permissively licensed models and developed in consultation with medical doctors.
% \item A dataset of 8,800 synthetic doctor-patient conversations with corresponding audio recordings and reference SOAP notes.
% \item An evaluation of current open-weight cascaded and end-to-end systems on the audio-to-SOAP-note task, finding that cascaded systems substantially outperform end-to-end approaches, with the cascaded pipeline achieving near-ceiling ASR performance.
% \end{itemize}
% We will release all data and code to support future work on long-context audio training and real-world validation.
% % All data and code will be made publicly available to support future work on long-context audio training and real-world validation.

\section{Related Work}

% \subsection{Long-Context Audio Reasoning and LALMs}

Recent advances have dramatically expanded the context windows of Large Audio Language Models (LALMs). While early attempts at end-to-end (E2E) speech summarization struggled with the quadratic memory complexity of processing long audio sequences~\cite{sharma2022end}, current systems can ingest continuous audio ranging from 40 minutes to over eight hours~\cite{xu2025qwen3omnitechnicalreport, goel2025audioflamingo3advancing}. However, the capacity to process long-form audio does not equate to the ability to reason over it. Recent evaluations, including MMAU-Pro~\cite{kumar2025mmauprochallengingcomprehensivebenchmark} and BLAB~\cite{ahia2025blabbrutallylongaudio}, reveal severe degradation on long-sequence multi-hop reasoning, a ``modality reasoning gap''~\cite{xiang2025closing} that manifests as representational drift and ``lexical dominance''~\cite{chen2025audiollmsreallylisten} when processing speech compared to text. E2E architectures still significantly trail cascaded ASR-to-text systems on complex open-ended generation tasks~\cite{billa2026cascadeequivalence}, a gap our dataset is designed to probe.

Automating clinical documentation, such as SOAP note generation, is a historically challenging task traditionally addressed through cascaded ASR-to-text systems. The primary bottleneck for advancing end-to-end multi-modal models in this domain is the severe scarcity of conversational data. While proprietary systems have leveraged thousands of hours of real clinical audio~\cite{shafran-etal-2020-medical,soltau21_interspeech}, strict privacy regulations like HIPAA prevent the distribution of these datasets to the open-source community~\cite{s25144305, wang-etal-2024-notechat}. 
Consequently, public benchmarks have been heavily restricted to small-scale datasets like PriMock57~\cite{papadopoulos-korfiatis-etal-2022-primock57}, which contains only 57 brief, actor-performed consultations that lack the organic complexity of genuine encounters, while ACI-Bench~\cite{yim2023acibench} provides 187 encounters in text form but no audio.
% Consequently, public audio-to-note benchmarks remain limited: PriMock57~\cite{papadopoulos-korfiatis-etal-2022-primock57} contains only 57 brief actor-performed consultations, and ACI-Bench~\cite{yim2023acibench}---the closest existing dataset to our task---provides 187 real-recorded encounters across three conversation types, far below the scale needed for training modern LALMs.

To circumvent these privacy barriers, recent research has successfully turned to Large Language Models (LLMs) to generate synthetic medical text. Multi-agent frameworks like NoteChat~\cite{wang-etal-2024-notechat} and single-prompt LLM simulators~\cite{s25144305} have demonstrated that models can produce highly realistic, medically accurate text transcripts of patient-physician interactions. However, these approaches remain strictly text-bound, failing to
address the acoustic realities of medical settings, e.g., multi-speaker
overlap, far-field microphone reverberation, and environmental
noise~\cite{soltau21_interspeech}. 
Our pipeline extends these text-only
approaches to full multi-speaker audio synthesis, advancing beyond
earlier work combining LLMs and TTS for conversational ASR~\cite{cornell24_syndata4genai} by integrating persona-conditioned voices and acoustic simulation to cross the significant ``sim2real'' gap of the clinical domain.
Given the inadequacy of surface metrics for clinical
summarization~\cite{liu-etal-2016-evaluate}, our evaluation adopts an
LLM-as-a-judge framework~\cite{zheng2023llmasajudge,liu2023geval}.

\section{Transcript and Audio Generation}

We now describe the three stages of our data generation pipeline, each targeting a specific gap identified above: (1) persona and context sampling, (2) persona-conditioned text dialogue generation, and (3) audio synthesis with acoustic simulation. Throughout, we use Gemma3-27B-IT~\cite{gemmateam2025gemma3technicalreport} as the LLM, selected for its performance in early dialogue generation tests, and SDialog~\cite{burdisso2026sdialogpythontoolkitendtoend} for pipeline orchestration.

\subsection{Sampling Personas}
Directly prompting an LLM to generate a first-visit conversation results in low diversity: out of 2,800 dialogues, we found only three unique doctor-patient pairs, with most sharing the same reason for visit and interpersonal dynamics. To increase diversity, we sample structured personas — lists of discrete attributes — for the doctor and patient before generation. We selected attributes that are few enough for the LLM to follow reliably, impactful enough to alter register or clinical content, minimally conflicting, and samplable from external demographic distributions to avoid internal LLM bias.

% Both the doctor and patient share the following attributes: name, age, height, weight, race, gender, forgetfulness, formality, hurriedness, and marital status. The patient has: fluency in English, occupation, insurance, and reason for visit. The doctor has years of experience. All attributes are sampled uniformly from predefined lists, using publicly available sources, except the reason for visit, which comes from a new source @Hass TODO. The persona generation stage produces 100 synthetic doctors and 200 synthetic patients.

Both the doctor and patient share the following attributes: name, age, height, weight, race, gender, forgetfulness, formality, hurriedness, and marital status. The patient has: fluency in English, occupation, insurance, and reason for visit. The doctor has years of experience. All attributes are sampled uniformly from predefined lists, using publicly available sources, except the reason for visit, which comprises 724 chief complaints compiled by a co-author with clinical expertise using an expert elicitation technique to collect a unique list of chief complaints across an entire healthcare system. These complaints cover typical and atypical primary care presentations that are clinical or administrative in nature (e.g., ``hand swelling,'' ``unexplained bruises on legs'', ``medical clearance'') and contain no personally identifiable information.
% These complaints represent typical reasons for visiting a primary care physician (e.g., "Hand Swelling," "Unexplained Bruises on Legs") and contain no personally identifiable information, as they are generic medical categories rather than data derived from patient records. The full list will be released along with the code used to generate the data. The persona generation stage produces 100 synthetic doctors and 200 synthetic patients.

\subsection{Personas $\rightarrow$ Text Dialogue}

We experimented with a direct generation approach, where the generator LLM receives both the doctor and patient personas as input and generates the entire dialogue in a single shot. We found that doing so led to unrealistically smooth dialogues, where the doctor quickly identified and handled the patient's reason for visit. We thus opted for a multi-turn generation approach, where the generator LLM produces a single turn of dialogue at a time for the doctor (resp. patient), and receives the dialogue history so far as well as the doctor's (resp. patient's) persona. 
On a subset of 2,800 dialogues, the multi-turn approach increased Claude Sonnet 4\footnote{dev evaluation only; not in the final pipeline} ratings of interpersonal challenge (1--5 scale) from $2.01$ (CI: $1.98$--$2.04$) to $2.50$ (CI: $2.46$--$2.54$). Persona adherence ratings likewise increased, from $3.10$ (CI: $3.07$--$3.13$) to $3.30$ (CI: $3.26$--$3.34$) (doctor) and from $3.60$ (CI: $3.58$--$3.62$) to $3.90$ (CI: $3.88$--$3.92$) (patient).
% On a subset of 2,800 dialogues, the multi-turn approach increased Claude Sonnet 4\footnote{dev evaluation only; not in the final pipeline} ratings of interpersonal challenge (1--5 scale) from $2.01 \pm 0.03$ to $2.5\pm 0.04$. Persona adherence ratings likewise increased, from $3.1 \pm 0.03$ to $3.3\pm 0.04$ (doctor) and from $3.6 \pm 0.02$ to $3.9\pm 0.02$ (patient). 
We hypothesize this is because the generator LLM can focus on the active role more specifically. Multiple rounds of physician feedback were collected on the text dialogue generation, leading to iterative refinement of the personas and prompts.

\begin{table}[t]
  \centering
  \caption{Dialogue statistics comparing our generated synthetic conversations against baseline and reference datasets.}
  \label{tab:transcript_stats}
  \footnotesize
  \setlength{\tabcolsep}{2.4pt}
  \begin{tabular}{l c c c c c}
    \toprule
    \multirow{2}{*}{\textbf{Dataset}} & \multicolumn{2}{c}{\textbf{Doctor}} & \multicolumn{2}{c}{\textbf{Patient}} & \multirow{2}{*}{\makecell{\textbf{Num}\\\textbf{Turns}}} \\
    \cmidrule(lr){2-3} \cmidrule(lr){4-5}
    & \makecell{\textbf{Turn}\\\textbf{Length}} & \makecell{\textbf{Fog}\\\textbf{Index}} & \makecell{\textbf{Turn}\\\textbf{Length}} & \makecell{\textbf{Fog}\\\textbf{Index}} & \\
    \midrule
    Ours         & $49.9 \pm 16.0$ & $10.4$ & $56.0 \pm 15.2$ & $6.9$ & $28.4 \pm 7.5$ \\
    PriMock57    & $18.8 \pm 4.7$ & $6.6$ & $12.3 \pm 5.8$ & $5.5$ & $97.3 \pm 17.7$ \\
    Mocks        & $29.6 \pm 6.3$ & $7.0$ & $13.6 \pm 3.1$ & $7.0$ & $54 \pm 6.2$ \\
    \bottomrule
  \end{tabular}
\end{table}

\begin{figure}[t]
  \centering
  \includegraphics[width=\linewidth]{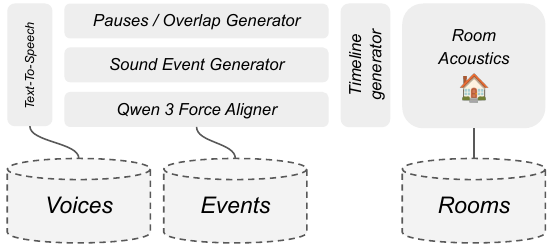}
  \caption{Overview of the audio generation pipeline, integrating Text-To-Speech, sound event generation, temporal scene composition, and environmental acoustic simulation.}
  % \caption{Overview of the audio generation pipeline, illustrating the integration of Text-To-Speech, sound event generation, and temporal scene composition with environmental acoustic simulation.}
  \label{fig:pipeline}
\end{figure}

% To assess whether the generated dialogues exhibit realistic conversational properties, we compare their turn length, Gunning fog index, and number of turns against two references (Tab.~\ref{tab:transcript_stats}): PriMock57~\cite{papadopoulos-korfiatis-etal-2022-primock57} and three in-person first-visit consultations recorded by two of the authors (including a licensed physician), which we refer to as Mocks.
% We see that the three transcript corpora are dramatically different. The pipeline-generated transcripts have fewer, longer turns, likely due to Gemma 3 being trained on primarily written, not spoken, language. Due to PriMock57's telehealth setting, it has more and shorter turns than our in-person Mocks. Our generated doctors exhibit greater linguistic complexity than the references.

% As Table~\ref{tab:transcript_stats} shows, our generated doctors exhibit higher language complexity (fog index 10.4 vs. 6.6–7.0) and longer turns than either reference set, likely because Gemma 3 was trained primarily on written text. We used direct prompting to encourage conversational style, emphasizing the in-person spoken setting, which had a moderate effect and is used in the final system. Few-shot examples from Mocks were also tested but caused personality and topic leakage without meaningfully reducing complexity, so they were excluded.

As Table~\ref{tab:transcript_stats} shows, our generated doctors exhibit higher language complexity (fog index 10.4 vs. 6.6–7.0) and longer turns than either reference set, likely because Gemma~3 was trained primarily on written text. PriMock57's telehealth setting further explains its higher turn count and shorter turn length relative to our in-person Mocks.\footnote{Three mock clinical encounter recordings (actor + real physician, realistic setting).} We used direct prompting to encourage conversational style, emphasizing the in-person spoken setting, which had a moderate effect and is used in the final system. Few-shot examples from Mocks were also tested, but they caused personality and topic leakage without meaningfully reducing complexity and were therefore excluded.

%%%%%%
% This can be hidden
%%%%%%
% \begin{figure}[t]
%   \centering
%   \includegraphics[width=\linewidth]{text_pipeline_A.pdf}
%   \caption{Text dialogue generation pipeline. Personas are uniformly
%     sampled from attribute lists and paired for multi-turn generation
%     with Gemma~3. Stage instructions are mapped to sound events for
%     audio synthesis.}
%   \label{fig:text_pipeline}
% \end{figure}

%%%%%%%%%%%%%%%%%%%%%%%%%%%%%%
%% AUDIO PIPELINE
%%%%%%%%%%%%%%%%%%%%%%%%%%%%%%
\subsection{Text Dialogue $\rightarrow$ Conversation Audio}

Using the generated personas (200 patients, 100 doctors), we synthesized 8,800 unique dialogues using a cross-product. The language models were prompted to incorporate natural conversational artifacts, such as colloquial speech patterns and implicit acoustic triggers (e.g., notations for door knocks or paper rustling). This design introduces a distinct ``cross-register'' modeling challenge: effectively mapping informal, spontaneous, and acoustically noisy spoken dialogue into formal, structured clinical documentation.

\subsubsection{Persona-Conditioned Voice Synthesis}
Translating these text dialogues into realistic audio requires a rigorous alignment of acoustic properties with the underlying linguistic personas. For each of the 300 unique personas, we perform conditioned voice cloning using the LibriTTS dataset~\cite{zen19_interspeech} samples of $30$s which have been normalized with peak normalization ($-1.0$ to $1.0$). This alignment utilizes gender, which is known in the dataset, to match dialogue personas. For age, an initial attempt was made to estimate speaker age ranges using the Qwen-3-Omni model~\cite{xu2025qwen3omnitechnicalreport}; however, a listening test revealed that the estimated ages were often inaccurate, particularly for elderly voices. Consequently, the initial random age assignment was retained. To maintain consistency and preserve the integrity of the training, development, and test subsets, this voice-to-persona matching is performed once and fixed for all experiments. Crucially, we ensure that the speakers designated for training, development, and testing are strictly disjoint.

\subsubsection{TTS Engine Selection and Subjective Assessment}

To establish a high-fidelity baseline, we considered nine TTS configurations (including \textit{Chatterbox}, \textit{Kokoro}, \textit{Qwen3}, \textit{XTTS}, and \textit{IndexTTS}) using LibriTTS profiles. \textit{XTTS} was excluded due to licensing restrictions. Qwen3-TTS-1.7B was selected for its balance of linguistic accuracy, expressiveness, and permissive licensing. The selected engine achieves a dry 
WER of $<2\%$ against the source text; to ensure consistent scoring, we utilize 
Whisper text normalization\footnote{\href{https://github.com/openai/whisper/blob/main/whisper/normalizers}{https://github.com/openai/whisper/blob/main/whisper/normalizers}} preceded by a custom ASCII normalization step (e.g., standardizing curly quotes) 
on both the textual utterance ground truths and the hypotheses.

\subsubsection{Acoustic Simulation and Flow Enrichment}

Real-world clinical encounters occur in complex acoustic environments; we enrich the TTS output through six pipeline stages (Fig.~\ref{fig:pipeline}): voice reference matching, neural speech synthesis, overlap \& pause insertion, sound event insertion, scene timeline composition using the dialogue variant\footnote{\href{https://github.com/dscaper/dscaper}{https://github.com/dscaper/dscaper}} of scaper~\cite{8170052}, and acoustic ray-tracing simulation with PyRoomacoustics~\cite{10.1109/ICASSP.2018.8461310}.

% The TTS output is convolved with a Room Impulse Response (RIR) for a typical $8m^2$ examination room. Additive background noise representative of clinical settings (e.g., HVAC hum) is then introduced. Finally, specific 66 discrete sound event classes (such as typing, paper rustling, or door knocks) are superimposed at predicted temporal offsets, using the Qwen3 forced alignment~\cite{shi2026qwen3asrtechnicalreport}, as dictated by the dialogue's acoustic triggers where the stage directions are described in the dialogue text and then mapped to corresponding audio event classes using Gemma~3. Finally, conversational dynamics, including natural overlaps and pauses, are computationally injected using a specialized Gemma3-27B-IT model prior to simulation to mimic genuine turn-taking behavior.

The TTS output is convolved with a Room Impulse Response (RIR) representative of a typical $8\text{m}^2$ examination room, followed by the addition of clinical background noise (e.g., HVAC hum). Subsequently, 66 discrete sound event classes, including typing, paper rustling, and door knocks, are superimposed at predicted temporal offsets. These offsets are determined via Qwen3 forced alignment~\cite{shi2026qwen3asrtechnicalreport} based on acoustic triggers within the dialogue's stage directions, which are mapped to audio event classes using Gemma~3. Finally, conversational dynamics, including natural overlaps and pauses, are computationally injected using Gemma 3 to mimic genuine turn-taking behavior.

% Evaluation across acoustic configurations reveals a nuanced trade-off between environmental reverberation and signal complexity. In 'Dry' settings, the model achieves its peak performance when handling overlaps and pauses, showing a 9\% relative improvement over the baseline configuration. While the most challenging 'Wet' conditions—combining both events and overlaps—lead to a relative increase in error of approximately 7\%, this shift is representative of the inherent difficulty of the task rather than a structural limitation. Furthermore, the model demonstrates notable resilience in 'Wet' configurations without overlaps; here, the inclusion of events yields a 4\% relative gain in accuracy, suggesting a robust synergy when processing complex acoustic artifacts in reverberant environments.

\subsubsection{Signal Augmentation}

To emulate realistic deployment constraints and acoustic conditions, we apply a comprehensive signal augmentation pipeline. We use the Opus codec at 16 kbps, yielding a compression ratio of approximately 14:1 %\footnote{7\% of the original WAV file size} 
and introducing realistic artifacts typical of real-world recordings. To enhance the perception of depth and compensate for the ray-tracing approach's limitations on the low end of the spectrum, the patient's audio amplitude is scaled down by a factor of 4 relative to the doctor's. Evaluation of the UTMOS metric~\cite{shi25f_interspeech, baba2024utmosv2} indicates that our generative framework achieves a score of 1.27, demonstrating a negligible performance gap when compared to our in-person Mocks baseline (1.28) and the DISPLACE-M~\cite{e2026benchmarkingspeechsystemsfrontline} challenge reference data (1.29). These results indicate that our synthetic output closely approximates the perceptual fidelity of authentic recordings.

\subsection{Summary of Dataset}
The Synth-DoPaCo (Synthetic Doctor Patient Conversations) dataset comprises 8,800 synthetic doctor-patient dialogues totaling 1,329 hours of audio. Each dialogue contains on average 28 turns ({\raise.17ex\hbox{$\scriptstyle\sim$}}14  per speaker), 1,500 words, and is augmented with approximately 37 non-speech audio events (e.g., coughs, background sounds) to simulate realistic clinical encounters. Dialogues average 9 minutes in length (range: 2--47\, min). The dataset is split into train, test, and development sets, as summarized in Table~\ref{tab:synth-dopaco-stats}.

\begin{table}[th]
  \caption{Synth-DoPaCo dataset statistics.}
  \label{tab:synth-dopaco-stats}
    % \small
  \footnotesize % Updated from \small to \footnotesize (approx 8pt)
  \centering
  \begin{tabular}{lrrrr}
    \toprule
    & \textbf{Train} & \textbf{Dev} & \textbf{Test} & \textbf{Overall} \\
    \midrule
    Personas (Doc/Pat) & 60/120 &  20/20 &  20/60 & 100/200 \\
    Dialogues            &  7,200 &    400 &  1,200 &   8,800 \\
    Hours              &  1,087 &     59 &    183 &   1,329 \\
    Words in dialogues &  10.9M &   582K &   1.8M &   13.3M \\
    Turns/dialogue       &   28.4 &   28.7 &   28.0 &    28.4 \\
    Duration/dialogue (s)&    544 &    530 &    548 &     544 \\
    Audio events/dialogue&   37.7 &   36.7 &   36.5 &    37.5 \\
    Words per SOAP note &   328.3 &    325.1 &    324.2 & 327.6 \\
    \bottomrule
  \end{tabular}
\end{table}
%\footnotetext{Preliminary estimate from a partial sample; generation ongoing at submission.}

% Add to Section 5, after the dataset statistics paragraph (after line 623):

On the wet\footnote{``wet'' final augmented audio, optionally Opus-compressed} audio, Whisper Large V3~\cite{radford2022robustspeechrecognitionlargescale} achieves 2--3\% WER while Qwen3-ASR~\cite{shi2026qwen3asrtechnicalreport} shows 10--14\%, confirming the audio is intelligible yet acoustically challenging. The Qwen3-ASR gap is primarily driven by increased substitution and deletion rates, exacerbated by Opus compression.

TTS synthesis was performed on a single NVIDIA A100~40\,GB GPU (approximately 2.5k GPU-hours); E2E or cascaded SOAP note generation using Qwen3-30B-A3B~\cite{yang2025qwen3technicalreport} ran on two NVIDIA A100~40\,GB GPUs via Ollama (approximately 300 GPU-hours).

\section{SOAP Note Generation and Evaluation}
Using the speaker-attributed transcript produced by the pipeline, we generate reference SOAP notes and evaluate all systems via two-stage processes, both using Kimi~K2~Thinking~\cite{kimiteam2026kimik2openagentic}. (Kimi~K2 inference was accessed via AWS Bedrock and Google Vertex and is not counted in GPU-hours.)

\subsection{Reference SOAP Note Generation}
We found that even large open-weight reasoning LLMs are prone to hallucinations in long-context text summarization tasks. In particular, (1) they tend to invent natural extensions of what occurred in the dialogue, e.g., a more detailed physical exam or follow-up plan, and (2) they use medical terms that are unsupported, e.g., describing a cold (which could be viral or bacterial) as a ``viral illness.'' To mitigate these issues, we enforce strict grounding in a fact-extraction stage that produces a structured JSON fact table, in which each fact is linked to a supporting quote and its turn index in the transcript. This fact table serves as the sole input to the note generation, preventing access to the transcript and limiting opportunities for hallucination. In the generation phase, we allow terminology to be rewritten into standard clinical language, but increasing clinical specificity or introducing new clinical content beyond the extracted facts is prohibited. 
A round of clinical feedback was integrated, including strict separation of the history of present illness (HPI) and review of systems (ROS), tighter integration of the assessment and plan, and prevention of over-documentation by excluding non-clinically relevant administrative content.

%To reduce these issues, a preliminary fact extraction stage is performed, which produces a structured JSON fact list, and this list is used as the sole input to the second, note generation, phase. A round of clinical feedback was integrated, reducing redundancy between the different sections of the SOAP note (and thus making it faster to read), making the assessment and plan more tightly integrated, and using medical terms when supported by evidence.

\subsection{SOAP Note Evaluation with LLM-as-a-Judge}
We evaluate SOAP notes using a two-stage LLM-as-a-judge pipeline~\cite{glover-etal-2022-revisiting}: atomic claims are first extracted from the note into a structured JSON representation, then each claim is assigned a support label indicating whether it is grounded in the transcript, with explicit evidence required for supported or contradicted claims.
We score SOAP notes along 12 dimensions.
Four dimensions are scored on a 1--5 scale (5 = best): \textbf{Faithfulness} (claim support and contradictions), \textbf{Structure} (SOAP formatting and section placement), \textbf{Coverage} (completeness of documentation relative to transcript evidence), and \textbf{Conciseness} (redundancy and low-value content). Six dimensions are reported as counts: Over-medicalization (unjustified medical interpretation), Under-medicalization (loss of explicitly stated medical specificity), Over-specificity (adds unjustified specificity), Missed relevant facts (missing transcript-grounded key facts), Critical omissions, and Duplicated content across sections. We additionally evaluate two rates: unsupported claim rate and contradiction rate. 

% \renewcommand{\labelitemii}{\raisebox{0.2ex}{\tiny$\bullet$}}
% \begin{itemize}
%     \item[]
%     \begin{itemize}
%     \item \textbf{Faithfulness} (claim support and contradictions)
%     \item \textbf{Structure} (SOAP formatting and section placement)
%     \item \textbf{Coverage} (completeness of documentation relative to transcript evidence)
%     \item \textbf{Conciseness} (redundancy and low-value content)
%     \end{itemize}
% \end{itemize}
% %faithfulness (claim support and contradictions), structure (SOAP formatting and placement), coverage (transcript-grounded checklist), and conciseness (redundancy and low-value content). 
% Six dimensions are reported as counts:
% \begin{itemize}
%     \item[]
%     \begin{itemize}
%         \item \textbf{Over-medicalization} (unjustified medical interpretation)
%         \item \textbf{Under-medicalization} (loss of explicitly stated medical specificity)
%         \item \textbf{Over-specificity} (adds unjustified specificity)
%         \item \textbf{Missed relevant facts} (missing transcript-grounded key facts)
%         \item \textbf{Critical omissions}
%         \item \textbf{Duplicated content across sections}
%     \end{itemize}
% \end{itemize}

%In the second, a structured JSON of evaluation scores is calculated. We found that the two-stage pipeline was more effective at identifying missed and hallucinated claims.

\subsection{Comparison of Open-Weight LALMs}
Using the generated conversation audio and SOAP notes, we compare the quality of the reference notes (using the reference-free LLM-as-a-judge pipeline) and the performance of current open-weight LALMs and cascaded ASR and LLM systems (Table~\ref{tab:judge}, showing a subset of judge dimensions).

We compare these references to four baseline systems. Two of these systems are cascaded, using Qwen3-ASR~\cite{shi2026qwen3asrtechnicalreport} or Whisper Large V3 to first process the audio into a transcript (which lacks speaker attributions) and then provide that transcript as input to Qwen3-30B-A3B~\cite{yang2025qwen3technicalreport}, with the instruction to generate a SOAP note. We compare these to Qwen3-Omni-Instruct and Qwen3-Omni-Thinking~\cite{xu2025qwen3omnitechnicalreport}, models that receive the audio conversation only and are instructed to produce a SOAP note directly.

First, we analyze the quality of the reference notes. They score highly using the reference-free LLM-as-a-judge and, in particular, are highly faithful to the transcripts (with an average faithfulness of 5/5). While coverage, structure, and conciseness are not perfect, they should act as strong training and evaluation signals for current LALMs. Second, we find that end-to-end and cascaded Qwen3 systems are similar in coverage and structure, but somewhat less faithful and concise. Third, end-to-end systems had per claim hallucination rates of 32\% vs. 22--24\% for cascaded vs. 1\% for reference, indicating that these tested systems are not deployable. 
%Both faithfulness and coverage achieve near-minimal scores for end-to-end approaches and show hallucination rates of around 60\% (vs.\ 20\% for cascaded and 1.6\% for reference), indicating widespread inability to process the facts of the transcript. 
%Structure ratings additionally suffer a $\sim$40\% decrease relative to their cascaded counterparts, indicating instruction following issues.

\begin{table}[th]
  \caption{Comparison of different note generation methods. Metrics: (Faith)fulness, (Cov)erage, (Struct)ure, and (Conc)iseness\label{tab:judge}}
  % \caption{Comparison of different note generation methods. Metrics: Faithfulness (Faith.), Coverage (Cov.), Structure (Struct.), and Conciseness (Conc.). \label{tab:judge}}
  
  \centering
  \footnotesize
  \setlength{\tabcolsep}{3pt}

  \begin{tabular}{l l c c c c}
    \toprule
    \multicolumn{2}{l}{\textbf{Architecture}} & \textbf{Faith.} & \textbf{Cov.} & \textbf{Struct.} & \textbf{Conc.}\\
    \midrule
    \multirow{3}{*}{Qwen3} & -ASR+Thinking   &
    {3.1{\scriptsize\,(±0.7)}} & {3.9{\scriptsize\,(±0.9)}} & {4.0{\scriptsize\,(±1.0)}} & {3.4{\scriptsize\,(±0.8)}} \\
                           & -Omni-Instruct  &
    {2.7{\scriptsize\,(±0.8)}} & {4.3{\scriptsize\,(±0.8)}} & {4.3{\scriptsize\,(±0.9)}} & {3.0{\scriptsize\,(±0.7)}} \\
                           & -Omni-Thinking  &
    {2.7{\scriptsize\,(±0.8)}} & {4.2{\scriptsize\,(±0.8)}} & {4.0{\scriptsize\,(±0.9)}} & {3.1{\scriptsize\,(±0.7)}} \\
    \multicolumn{2}{l}{Whisper Large V3+}    &
    \multirow{2}{*}{3.3{\scriptsize\,(±0.8)}} &
    \multirow{2}{*}{4.3{\scriptsize\,(±0.8)}} &
    \multirow{2}{*}{4.0{\scriptsize\,(±0.9)}} &
    \multirow{2}{*}{3.5{\scriptsize\,(±0.7)}} \\
    \multicolumn{2}{l}{Qwen3-Thinking}       &  &  &  &  \\
    \midrule
    \multicolumn{2}{l}{Reference (Oracle}    &
    \multirow{2}{*}{5.0{\scriptsize\,(±0.2)}} &
    \multirow{2}{*}{4.2{\scriptsize\,(±0.9)}} &
    \multirow{2}{*}{4.5{\scriptsize\,(±0.9)}} &
    \multirow{2}{*}{3.8{\scriptsize\,(±0.8)}} \\
    \multicolumn{2}{l}{+ Kimi K2 Thinking)}  &  &  &  &  \\
    \bottomrule
  \end{tabular}
\end{table}

\subsection{Reference-Grounded SOAP Note Evaluation}
In addition to the reference-free LLM judge, we evaluate generated notes against the reference using standard ROUGE~\cite{rouge-score} F1 metrics (R-2, R-3, R-L) and an Open Medical Concept metric\footnote{Medical concept F1: MeSH keyword matching + NER via \texttt{en\_core\_sci\_md} (\texttt{scispaCy}~\cite{neumann-etal-2019-scispacy}).} inspired by \cite{zhang-etal-2021-leveraging-pretrained}, which extracts medical concepts from both reference and hypothesis notes and computes F1 over their overlap (Table~\ref{tab:comparison-DD-400}).
%These metrics also show a large gap between cascaded and end-to-end performance.
On these, we find comparable or higher scores for end-to-end systems.

\begin{table}[htbp]
\centering
\caption{Dev-set SOAP-note F1 (\%); $\pm$ is the bootstrapped 95\% CI half-width.}
\label{tab:comparison-DD-400}
  \footnotesize
\setlength{\tabcolsep}{4pt}
{\fontsize{9pt}{11pt}\selectfont
\renewcommand{\arraystretch}{1.3}
\begin{tabular}{l c c c c r}
\toprule
Model & {R-2} & {R-3} & {R-L} & {Open} & {\#Wrd} \\
\midrule
\makecell[l]{Whisper+Qwen3}       & \ci{11.8}{0.3} & \ci{4.16}{0.20} & \ci{22.7}{0.4} & \ci{29.1}{0.5} & \makecell[r]{255} \\
\addlinespace[6pt]
\makecell[l]{Qwen3-ASR+Qwen3}     & \ci{11.0}{0.3} & \ci{3.84}{0.22} & \ci{21.9}{0.5} & \ci{27.9}{0.6} & \makecell[r]{254} \\
\midrule
\makecell[l]{Qwen3-Omni-Thinking} & \ci{12.7}{0.3} & \ci{5.03}{0.21} & \ci{21.3}{0.4} & \ci{27.5}{0.5} & \makecell[r]{487} \\
\addlinespace[6pt]
\makecell[l]{Qwen3-Omni-Instruct} & \ci{14.3}{0.4} & \ci{6.24}{0.27} & \ci{22.9}{0.4} & \ci{30.9}{0.6} & \makecell[r]{545} \\
\bottomrule
\end{tabular}}
\end{table}

Several limitations warrant consideration. First, the reference SOAP notes are LLM-generated rather than authored by physicians, which may introduce systematic biases in both the content and the evaluation signal. Second, all dialogues are in English, two-speaker, primary-care first-visit encounters, limiting generalizability to other languages, clinical specialties, or multi-party settings. Third, the low WER achieved on wet audio ($<$3\% for Whisper) suggests that the acoustic simulation may not fully replicate the difficulty of real clinical recordings, and a direct sim-to-real comparison remains for future work.

\section{Conclusion}

We presented a fully synthetic pipeline---persona-conditioned dialogue generation, multi-speaker audio synthesis, and reference SOAP note production---built entirely on open-weight models, yielding 8,800 conversations and 1,329 hours of audio. End-to-end and cascaded systems perform comparably, with cascaded systems likely being preferred in practice due to their higher faithfulness and lower hallucination rate. With Whisper Large V3 achieving 2--3\% WER on wet audio, ASR is near ceiling; the primary challenge lies not in transcription but in reasoning over long, noisy conversations to produce faithful clinical documentation.
Future work aims to close the gap between synthetic and real clinical audio — through multi-party scenarios (e.g., nurse, caregiver), accent-conditioned voices, and more natural conversational patterns — and to extend the pipeline to other professional domains. 
Data and code will be publicly released and accessible from Hugging Face\footnote{\href{https://huggingface.co/datasets/Play-Your-Part/Synth-DoPaCo}{https://huggingface.co/datasets/Play-Your-Part/Synth-DoPaCo}}. The test data will be withheld until December 2026.

\section{Acknowledgement}
% \section*{Acknowledgments}
The authors would like to thank Markus Müller (Amazon AGI) for his valuable discussions, leadership, and guidance throughout the duration of the workshop. The contribution by Markus Müller was made in his capacity as a workshop leader and is not related to his position at Amazon. %does not necessarily reflect the official position of Amazon. 

This work was supported by the 2025 Jelinek Memorial Summer Workshop on Speech and Language Technologies (JSALT 2025), hosted at the Brno University of Technology and organized by the Center for Language and Speech Processing (CLSP) at Johns Hopkins University.
% \textcolor{red}{
We gratefully acknowledge the Ohio Supercomputer Center (OSC) for providing computational resources used in this work~\cite{OhioSupercomputerCenter1987}.
The Idiap employees were funded by European Union Horizon 2020 project ELOQUENCE (101070558). Severin Baroudi and Ricard Marxer benefited from the support of the Agence de l’Innovation Defense under the ``grant number 2022 65 0079'', the French National Research Agency through the ANR-20-CE23-0012-01 (MIM) grant, and computational resources provided by GENCI-IDRIS HPC (Grant AD011014044R2).
% }
% \textcolor{red}{
% The development of the audio module was performed partially using HPC resources from GENCI-IDRIS (Grant AD011013061R3) and was financially supported by ANR MALADES (ANR-23-IAS1-0005) and BPI PARTAGES.
% }
% \textcolor{red}{
% XXXX benefited from the support of the French National Research Agency through the ANR-20-CE23-0012-01 (MIM) grant, supported by the Agence de l’Innovation Defense under the ``grant number 2022 65 0079'' and computational ressources provided by GENCI-IDRIS HPC (Grant AD011014044R2).
% }

\section{Generative AI Use Disclosure}

Generative AI tools were used in two distinct ways in this work.

\textbf{Manuscript preparation.}
Large language models were used to assist with proofreading, improving conciseness, and formatting \LaTeX{} tables. All such use was directed and reviewed by an author; AI tools produced no significant portions of the manuscript without subsequent human revision. AI assistance was also used in the development of experimental code and scripts for running experiments on HPC clusters. Claude Sonnet 4\footnote{AWS Bedrock model\\ \texttt{global.anthropic.claude-sonnet-4-20250514-v1:0}} was used for development-only evaluation of the text dialogue generation pipeline and is not a component of the released dataset or final system.

\textbf{Research methodology.}
Generative AI models are integral components of the proposed pipeline and are described in full detail in the body of the paper. Specifically: Gemma3-27B-IT\footnote{https://huggingface.co/google/gemma-3-27b-it}~\cite{gemmateam2025gemma3technicalreport} was used to generate all synthetic dialogues; Qwen3-TTS~1.7B~\cite{yang2025qwen3technicalreport} was used for speech synthesis; and Kimi~K2~Thinking\footnote{AWS Bedrock model \texttt{kimi.moonshot.k2-thinking-v1:0}}~\cite{kimiteam2026kimik2openagentic} was used both for reference SOAP note generation and as the LLM-as-a-judge evaluator. Additionally, Qwen3-ASR\footnote{https://huggingface.co/Qwen/Qwen3-ASR-1.7B}~\cite{shi2026qwen3asrtechnicalreport}, Qwen3-30B-A3B~\cite{yang2025qwen3technicalreport}, and Qwen3-Omni\footnote{https://huggingface.co/Qwen/Qwen3-Omni-30B-A3B-Thinking and https://huggingface.co/Qwen/Qwen3-Omni-30B-A3B-Instruct}~\cite{xu2025qwen3omnitechnicalreport} were evaluated as baseline systems and are subjects of investigation in this work. These uses constitute the research contribution of the paper and are not uses of AI for manuscript authorship.

All co-authors have reviewed and take responsibility for the full content of this paper.

%\section{References}
% BibTeX entries go here.
\bibliographystyle{IEEEtran}
\bibliography{refernces}

\end{document}